\documentclass[format=sigconf]{acmart}

\usepackage{booktabs} 
\setcopyright{rightsretained}



\copyrightyear{2018}
\acmYear{2018}
\setcopyright{acmlicensed}
\acmConference[ESEM '18]{ACM / IEEE International Symposium on Empirical Software Engineering and Measurement (ESEM)}{October 11--12, 2018}{Oulu, Finland}
\acmBooktitle{ACM / IEEE International Symposium on Empirical Software Engineering and Measurement (ESEM) (ESEM '18), October 11--12, 2018, Oulu, Finland}
\acmPrice{15.00}
\acmDOI{10.1145/3239235.3267434}
\acmISBN{978-1-4503-5823-1/18/10}

\usepackage[disable]{todonotes}

\usepackage[utf8]{inputenc}
\usepackage{url}
\usepackage{enumitem}
\usepackage{flushend}
\usepackage{graphicx}
\usepackage{amssymb}

\usepackage{subcaption}
\usepackage{xspace}
\usepackage{hyperref}

\newcommand{\fig}[1]{Fig.~\ref{#1}}
\newcommand{\tab}[1]{Table~\ref{#1}}

\begin{document}
%

\title{On the Use of Emoticons in Open Source Software Development}


\author{Maëlick Claes}
\affiliation{M3S, ITEE, University of Oulu, Finland}
\email{maelick.claes@oulu.fi}

\author{Mika Mäntylä}
\affiliation{M3S, ITEE, University of Oulu, Finland}
\email{mika.mantyla@oulu.fi}

\author{Umar Farooq}
\affiliation{M3S, ITEE, University of Oulu, Finland}
\email{umar.farooq@oulu.fi}

\begin{abstract}

\textbf{Background}: Using sentiment analysis to study software developers' behavior comes with challenges such as the presence of a large amount of technical discussion unlikely to express any positive or negative sentiment. However, emoticons provide information about developer sentiments that can easily be extracted from software repositories.
\textbf{Aim}: We investigate how software developers use emoticons differently in issue
trackers in order to better understand the differences between developers and determine to which extent emoticons can be used as in place of sentiment analysis.
\textbf{Method}: We extract emoticons from 1.3M comments from Apache's issue tracker and 4.5M from Mozilla's issue tracker using regular expressions built from a list of emoticons used by SentiStrength and Wikipedia. We check for statistical differences using Mann-Whitney U tests and determine the effect size with Cliff's $\delta$.
\textbf{Results}: Overall Mozilla developers rely more on emoticons
than Apache developers. While the overall ratio of comments with
emoticons is of 2\% and 3.6\% for Apache and Mozilla, some individual
developers can have a ratio above 20\%. Looking specifically at Mozilla
developers, we find that western developers use significantly more
emoticons (with large size effect) than eastern developers. While the majority of emoticons are used to
express joy, we find that Mozilla developers use emoticons more
frequently to express sadness and surprise than Apache developers. Finally, we find that developers use overall more emoticons during weekends than during weekdays, with the share of sad and surprised emoticons increasing during weekends.
\textbf{Conclusions}: While emoticons are primarily used to express joy, the more occasional use of sad and surprised emoticons can potentially be utilized to detect frustration in place of sentiment analysis among developers using emoticons frequently enough.
\end{abstract}

\maketitle


\section{Introduction}

Decentralized online software development has grown during the past
decades to a point where thousands of developers collaborate on a
daily basis using electronic communication means. In software
engineering, many empirical studies have studied developers
interactions and behavior with mining repository techniques.

In particular, sentiment analysis has become more and more popular in
the past years to study developers' behavior at a large scale. Because
most gold standards for sentiment analysis are not based on software
engineering data~\cite{Novielli2015}, generic sentiment analysis tools
can often perform poorly~\cite{jongeling2015choosing} for analyzing
software development data. Thus, software engineering specific gold
standards for sentiment analysis have been
developed~\cite{islam2017leveraging, calefato2018sentiment}.  Although
such approaches improve sentiment analysis tool performance it may not
be enough~\cite{lin2018sentiment}. Additionally, they require tedious
manual labeling. Emoticon usage on large software engineering text
masses may lead to automated sentiment labeling and dictionary
building \cite{felbo2017using}.

One of the challenge with sentiment analysis in a software engineering
context is the large amount of neutral sentences due to the technical
nature of discussion. Moreover, it can also be difficult for human
beings~\cite{imtiaz2018sentiment} to interpret emotions based solely
on text. Emoticons are a popular way to add information about
expressed emotion in online textual conversations and can be used to
improve sentiment analysis tools~\cite{boia2013worth}. However,
emoticon usage can vary between persons, culture or from one context
to another~\cite{park2013emoticon}.

In this paper, we present the early results of an investigation of the
differences in usage of emoticons in open source projects from Mozilla
and Apache. Our goal is to understand how developers use emoticons
differently and to which extent emoticons can be used as a proxy to
measure developer behaviors. We define the following research
questions:

\begin{itemize}
\item How does the use of emoticons vary among developers?
\item What kind of emotions are expressed with emoticons?
\item How does the use of emoticons vary across a week?
\end{itemize}

The remaining of this paper is structured as follows. In section 2 we
present the related work. In section 3 we present the methodology used
to collect data, and how we extracted emoticons from issue
comments. In section 4 we present preliminary results for each of our
research questions. In section 5, we present the limits of our study
and we finally conclude in section 6.



\section{Related Work}\label{sec:related}

In the last two decades, emoticons have been widely used for communication on the Internet and considered as the paralanguage of the Web~\cite{marvin1995spoof}. As identified in ~\cite{rezabek1998visual}, 6.1\% of electronic mails contains emoticons and 13.2\% of the posts on UseNet newsgroup have emoticons~\cite{witmer1997line}.
Emoticons are exploited in both lexicon based and machine learning approaches of sentiment analysis to improve the accuracy of existing lexicon dictionary based sentiment analysis~\cite{hogenboom2013exploiting, bahri2018novel, novak2015sentiment, hogenboom2015exploiting, liu2012emoticon, davidov2010enhanced}.
As shown in~\cite{wang2015sentiment} the prediction capability of tweets as positive and negative is considerably enhanced when the emoticons are considered while training the classifier. One of the issue with traditional machine learning technique for sentiment classification is the dependency on topic, domain or even on natural language. An effort is made in~\cite{read2005using} in order to reduce this dependency by training the classifier using an emoticon corpus.

In most literature, emoticons are investigated in social media
contents. Differences in emoticon usage have been found between
gender, showing that in a same sex context, females use more emoticons
than males, while in a mixed sex context, male tend to adopt the same
standard of expression~\cite{wolf2000emotional}. Park et
al.~\cite{park2013emoticon} showed that emoticon usage differs between
geographical area and social network. Moreover, they are generally
positive and strongly negative sentences generally don't contain
emoticons. Miller et al.~\cite{miller2016blissfully} pointed out that
there can be disagreement among users about emojis being positive,
negative or neutral.

However, in this study, we are investigating it in more formal and technical communication of developers during software development. According to the best of our knowledge, this is the first effort to study the emoticons in software engineering in order to primarily focus on how developers use emoticons and which types of emoticons are more frequent.




\section{Methodology}\label{sec:methodo}

In this section, we present first the methodology used for extracting
data from Apache's and Mozilla's issue trackers. Then we give details on
how we identified emoticons from the data alongside some preliminary numbers
about the presence of emoticons in Mozilla and Apache comments.

\subsection{Data extraction}

To answer our research questions, we mined data from
Mozilla's\footnote{\url{https://bugzilla.mozilla.org}} and
Apache's\footnote{\url{https://issues.apache.org/jira/}} issue
trackers. We specifically extracted the history of all comments with
associated authors, projects and timestamps using \emph{Perceval} from
the \emph{GrimoireLab} tool chain~\cite{DuenasPerceval}. Because issue
tracker comments don't contain the timezone of the commenter, we also
extracted information with \emph{Perceval} about commits from
Mozilla's Mercurial
repositories\footnote{\url{https://hg.mozilla.org}} and Apache's Git
repositories\footnote{\url{https://git.apache.org/}}. In order to link
developer timezones from the commit repositories to their isssue
tracker comments, we performed a basic merging of the different
authors' identities. We grouped together identities using the same
name or email addresses after cleaning. Two of the authors manually
checked the result in order to avoid any false positive. We only kept
the (merged) developers who had made at least 100 commits. We end up
with 102 Apache projects and 66 Mozilla projects with at least 1000
comments. In total, these amounts to 1.3M Apache comments and 4.5M
Mozilla comments.



\subsection{Identifying emoticons}

In order to detect emoticons, we first used
NLoN~\cite{mantyla2018nlon} to discard lines of text that are not
natural language, such as code and execution traces. We used the list
of ASCII emoticons used by SentiStrength and completed it with common
ASCII emoticons from
Wikipedia\footnote{\url{https://en.wikipedia.org/wiki/List_of_emoticons}}. From
this list, we built regular expressions in order to extract the
emoticons present in comments. We found 84 of the 107 emoticons from
our list that were used at least once in a comment. We manually
classified the emoticons based on basic emotions defined by Plutchik~\cite{plutchik1991emotions}: joy (42
emoticons), sadness (11), surprise (25), anger (4) and unknown (2). We
made our list of emoticons and associated emotions available at
\url{https://github.com/M3SOulu/ESEM2018-Emoticons-Emotions-List}.

\begin{table}[ht]
  \centering
  \caption{Number of authors, comments and emoticons, percentage of
    comments with at least one emoticon, and percentage of authors having
    used at least one emoticon for Apache and Mozilla.}
  \label{tab:emoticons-by-source}
  \begin{small}
    \begin{tabular}{l|r|r|r}
      source & comments & authors & emoticons \\
      \hline
      Apache & 1,262,833 (2.12\%) & 956 (68.83\%) & 27,903 \\
      Mozilla & 4,466,817 (3.65\%) & 472 (97.46\%) & 172,868 \\
    \end{tabular}
  \end{small}
\end{table}

\tab{tab:emoticons-by-source} shows for Apache and Mozilla the number
of unique developers found in the issue trackers and the percentage
who has used at least one emoticon, the total number of comments and
percentage of comments with at least one emoticon, and the total
number of emoticons detected. Overall most comments only contain at
most one emoticon. Only 3.7\% of Apache comments with emoticons and
4.9\% of Mozilla comments with emoticons contain 2 emoticons. Less
than 1\% of the comments with emoticons contain more than 2 emoticons.



\section{Empirical analysis}\label{sec:analysis}





\subsection{Variation among developers}

\begin{figure}[!htp]
  \centering
  \includegraphics[width=0.75\columnwidth]{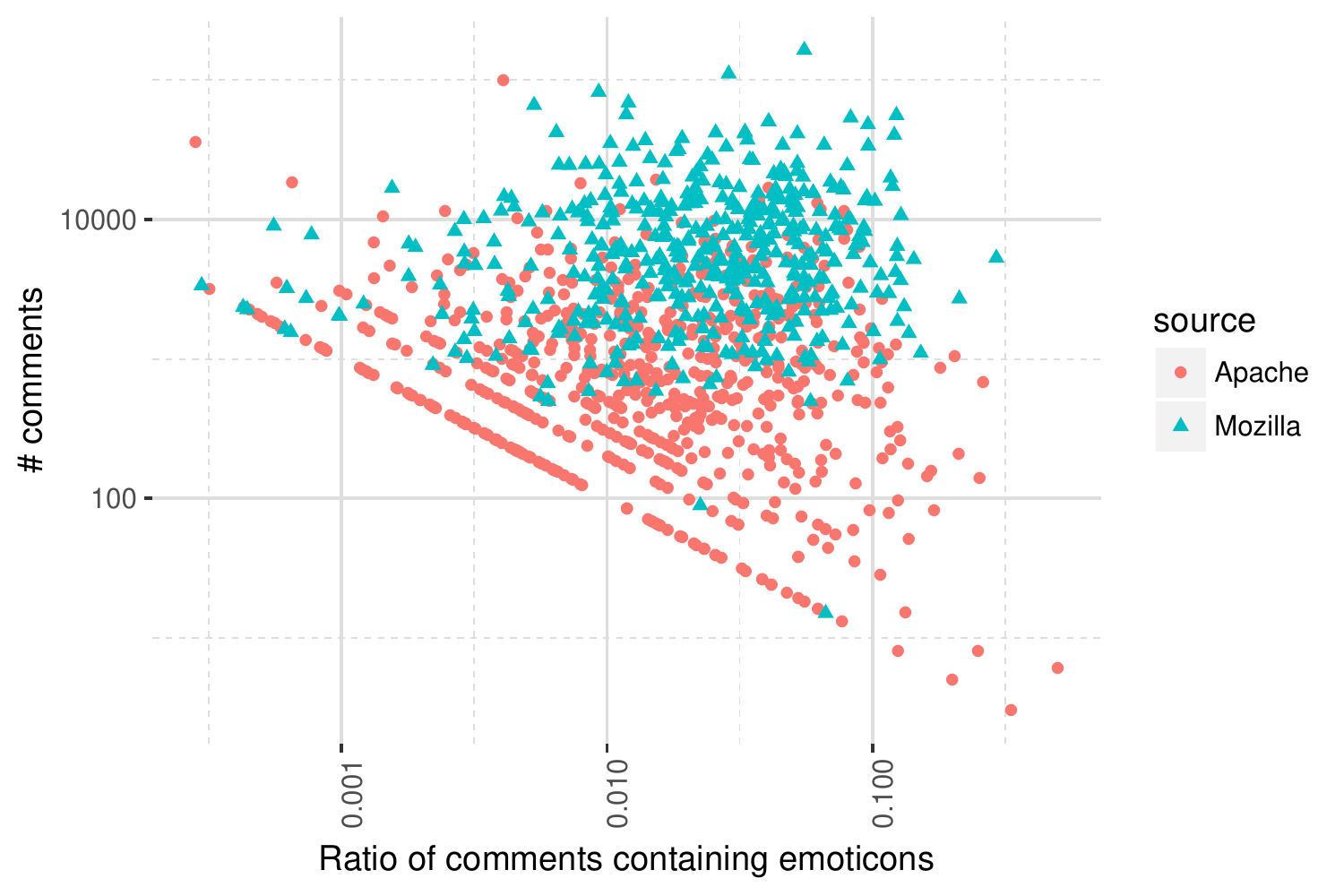}
  \caption{Number of comments and ratio of comments with at least one
    emoticon for each developer.}
  \label{fig:authors-ncomments-comment-ratio}
\end{figure}

As seen in \fig{fig:authors-ncomments-comment-ratio}, there are some
important differences in the frequency of emoticon usage between both
Apache and Mozilla developers. However, we find that there are also
Apache developers who have used emoticons in more than 10\% of their
comments.

Mozilla developers are usually more active on their issue tracker than
Apache developers. 
The difference
can be explained by the fact that Mozilla forces developers to use the
issue tracker for code review. On the other hand, Apache gives more
freedom to individual projects to handle modifications and code
review. For the remaining of this section, we only keep the developers
having written at least 100 comments. This leaves 727 Apache
developers (76\%) and 462 (98\%) Mozilla developers, having written
respectively 99.3\% and 99.9\% of all the comments.
Even after filtering out the developers with less
than 100 comments, there is still a statistically significant (p-value
$<$ 0.001) difference between Apache and Mozilla developers with a
medium effect size (Cliff's $\delta = 0.39$). After filtering out the
developers who haven't used any
emoticon, 
Mozilla developers still use emoticons significantly more often than
Apache developers with a small effect size (Cliff's $\delta = 0.27$).



\begin{figure}[!htp]
  \centering
  \includegraphics[width=0.75\columnwidth]{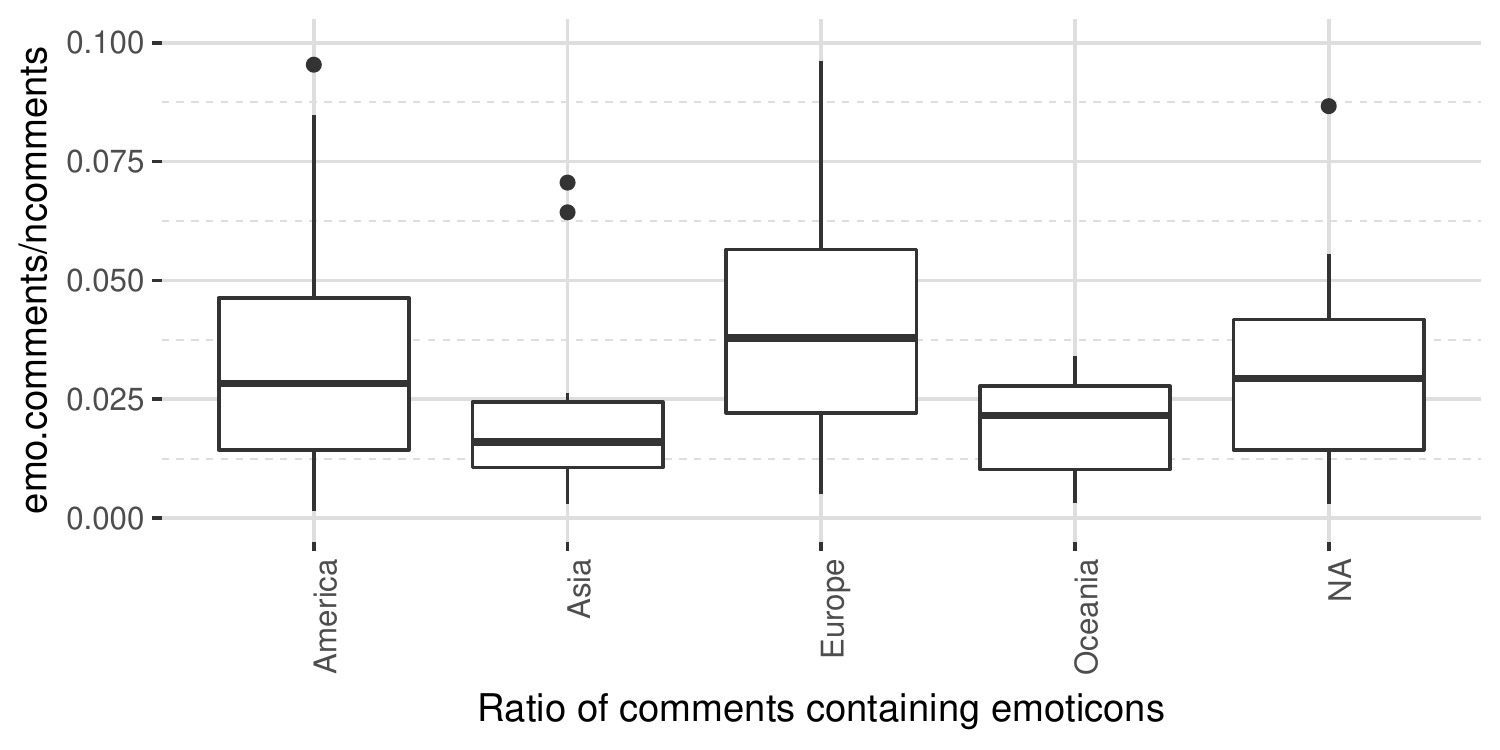}
  \caption{Ratios of comments with at least emoticons for hired
    Mozilla developers based on continents.}
  \label{fig:boxplot-mozilla-continent}
\end{figure}

Specifically for Mozilla, we manually extracted information about developers that were hired by the Mozilla Corporation in order to find out if we could identify demographic characteristics that can explain the differences of emoticon usage between developers. We didn't find any statistical differences between hired and volunteer developers. Moreover, for hired developers we also didn't find any statistically significant difference in terms of gender, position (managers vs. non manager, senior vs. non senior) or experience. We only observe a significant difference (p-value $=$ 0.002) between developers based in Europe or North America, and developers in Asia (Taiwan and Japan) or Oceania (Australia and New Zealand) with a medium size effect (Cliff's $\delta = -0.41$). As seen in \fig{fig:boxplot-mozilla-continent}, western developers are more likely to use emoticons than eastern developers.

\subsection{Emotions expressed}

In accordance with the literature, the majority of emoticons are used to express joy. As seen in \tab{tab:emotions-by-source}, besides joy, emoticons are mostly used to express sadness and surprise. We also looked at anger emoticons but these were used less than 1\% of the time. Differences exist between the Apache and Mozilla communities. In Apache, more than 90\% of the emoticons express joy. In Mozilla where emoticons are more frequently used than in Apache, sad (15\%) and surprised (7\%) emoticons are also used more frequently than in Apache (7\% and 2\%).

\begin{table}[ht]
  \centering
  \caption{Percentage of emoticons used to express joy, sadness and surprise.}
  \label{tab:emotions-by-source}
  \begin{small}
    \begin{tabular}{l|rrrr}
      source & \% joy & \% sadness & \% surprise & \% other \\
      \hline
      Apache & 90.13 & 7.27 & 2.3 & 0.34 \\
      Mozilla & 78.29 & 14.93 & 6.73 & 0.05 \\
    \end{tabular}
  \end{small}
\end{table}

There are also differences in term of emotions expressed between developers. 50.3\% of Apache developers, who have used emoticons, have only used them to express joy, while only 7.4\% of the Mozilla developers have.
28.4\% of the Mozilla developers have used more than 90\% of emoticons to express joy and 73.5\% of the Apache. On the other extreme, 6.1\% of the Mozilla developers and 5.6\% Apache developers have used at least as many sad and surprised emoticons as joyful ones.

\subsection{Variation across the time of the week}

\begin{table}[ht]
  \centering
  \caption{Statistical difference (p-vlaue and Cliff's $\delta$ of Mann-Whitney U test) between
    the ratio of issue comments with emoticons during the weekend and
    during the week.}
  \label{tab:weekend-variations}
  \begin{small}
    \begin{tabular}{l|r|r|r|r}
      Source & Emoticons & Joy & Sadness & Surprised \\
      \hline
      Mozilla & 0.17 (0.04) & 0.08 (-0.05) & $<$ 0.001 (0.15) & $<$ 0.001 (0.12) \\
      Apache & $<$ 0.001 (0.22) & $<$ 0.001 (-0.19) & $<$ 0.001 (0.18) & $<$ 0.001 (0.11)
    \end{tabular}
  \end{small}
\end{table}

\begin{figure}[!htp]
  \centering
  \includegraphics[width=0.75\columnwidth]{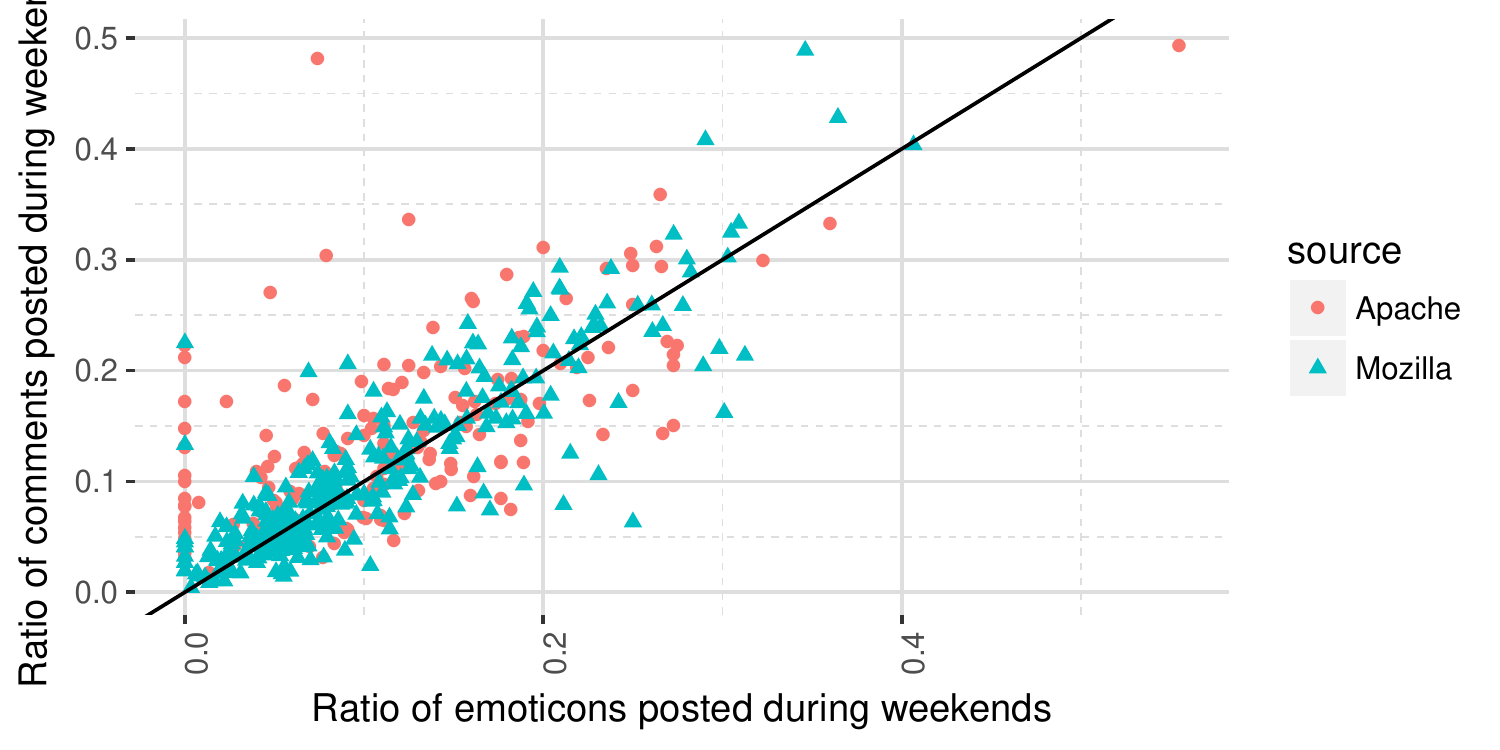}
  \caption{Developers' ratio of comments during weekends and ratio of
    emoticons during weekends.}
  \label{fig:weekend-ratios}
\end{figure}

\tab{tab:weekend-variations} shows the p-value and effect sizes of a Mann-Whitney U test between the ratio of developers' comments with emoticons during the weekend and during weekdays. It is for Apache where the differences are more statistically significant and have stronger effect sizes.
Emoticons are used more often in Apache projects during the weekend than during weekdays. \fig{fig:weekend-ratios} shows that this difference is observable in Apache because there are a number of developers with a relatively small percentage of comments made during weekends (5 to 20\%) but a large percentage of emoticons used during weekends ($> 25$\%). On the other hand, weekends have a lower proportion of joyful emoticons and high proportions of sad and surprised emoticons. This difference is observed in Apache, and in Mozilla projects only for sad and surprised emoticons.

\section{Threats to Validity}\label{sec:threats}


We merged developers' identities using a very basic
technique. Although we manually checked for false positives, there
might be false negatives remaining.
We only identified the most common ASCII emoticons and might have
missed less common emoticons. Moreover, we manually classified
emotions associated with emoticons and only associated one emotion to
each emoticon. Differences of context (irony, sarcasm or different
interpretation by the users) could imply a different meaning or
emotion for some emoticons.
Our study only includes open source projects from Apache and
Mozilla. The results obtained are specific to those organization's
culture and developer habits (external validity). Thus we cannot
guarantee that our data set would be representative of the entire
software industry or even open source industry.


\section{Conclusion and future work}\label{sec:conclusion}

In this paper, we conducted a preliminary investigation of the use of emoticons by software developers in Apache and Mozilla issue trackers. First, most Mozilla developers have used at least one emoticon, while more than 30\% of the Apache developers never used any. There are also differences between individual developers and individual projects.
We looked specifically at Mozilla's developers but couldn't find any statistical difference showing that professional developers might use emoticons differently than volunteers. Looking specifically at Mozilla's hired developers, we found only a difference between developers based in North America or Europe, and Asia or Oceania. The former using emoticons more frequently than the latter with a medium effect size.
These differences between developers show that emoticons are most likely to act as a good proxy for sentiment analysis for developers frequently relying on emoticons.

In accordance to previous work, Mozilla and Apache developers mostly use emoticons to express joy. More than 90\% of the emoticons used by Apache developers express joy and 78\% for Mozilla developers. Most other emoticons are used to indicate either sadness (7\% for Apache and 15\% for Mozilla) or surprise (2\% for Apache and 7\% for Mozilla). The presence of a significant amount of sad and surprised emoticons show that it is potentially feasible to partly detect negative emotions such as frustration of developers without advanced sentiment analysis. We found a small difference in emoticon usage during the weekend and weekdays. During weekends, developers use more emoticons but these are less positives and include more sad or surprised emoticons.


As this paper is only a first step towards understanding how emoticons are used in software development, we will extend our research in different directions. First we want to study the context in which software developers are more likely to use emoticons. For this, we plan on conducting a survey and ask directly to software developers how they use emoticons. Second, we want to compare emoticon usage with the result of sentiment analysis tools.
Finally we also intend find correlate emoticons usage with sentiment polarity, such as valence and arousal.



\section*{Acknowledgments}

The authors have been supported by Academy of
Finland grant 298020.





\bibliographystyle{IEEEtran}

\bibliography{biblio}

\providecommand{\noopsort}[1]{}
\begin{thebibliography}{10}
\providecommand{\url}[1]{#1}
\csname url@samestyle\endcsname
\providecommand{\newblock}{\relax}
\providecommand{\bibinfo}[2]{#2}
\providecommand{\BIBentrySTDinterwordspacing}{\spaceskip=0pt\relax}
\providecommand{\BIBentryALTinterwordstretchfactor}{4}
\providecommand{\BIBentryALTinterwordspacing}{\spaceskip=\fontdimen2\font plus
\BIBentryALTinterwordstretchfactor\fontdimen3\font minus
  \fontdimen4\font\relax}
\providecommand{\BIBforeignlanguage}[2]{{%
\expandafter\ifx\csname l@#1\endcsname\relax
\typeout{** WARNING: IEEEtran.bst: No hyphenation pattern has been}%
\typeout{** loaded for the language `#1'. Using the pattern for}%
\typeout{** the default language instead.}%
\else
\language=\csname l@#1\endcsname
\fi
#2}}
\providecommand{\BIBdecl}{\relax}
\BIBdecl

\bibitem{Novielli2015}
\BIBentryALTinterwordspacing
N.~Novielli, F.~Calefato, and F.~Lanubile, ``The challenges of sentiment
  detection in the social programmer ecosystem,'' in \emph{Proceedings of the
  7th International Workshop on Social Software Engineering}, ser. SSE
  2015.\hskip 1em plus 0.5em minus 0.4em\relax New York, NY, USA: ACM, 2015,
  pp. 33--40. [Online]. Available:
  \url{http://doi.acm.org/10.1145/2804381.2804387}
\BIBentrySTDinterwordspacing

\bibitem{jongeling2015choosing}
R.~Jongeling, S.~Datta, and A.~Serebrenik, ``Choosing your weapons: On
  sentiment analysis tools for software engineering research,'' in
  \emph{Software maintenance and evolution (ICSME), 2015 IEEE international
  conference on}.\hskip 1em plus 0.5em minus 0.4em\relax IEEE, 2015, pp.
  531--535.

\bibitem{islam2017leveraging}
M.~R. Islam and M.~F. Zibran, ``Leveraging automated sentiment analysis in
  software engineering,'' in \emph{Proceedings of the 14th International
  Conference on Mining Software Repositories}.\hskip 1em plus 0.5em minus
  0.4em\relax IEEE Press, 2017, pp. 203--214.

\bibitem{calefato2018sentiment}
F.~Calefato, F.~Lanubile, F.~Maiorano, and N.~Novielli, ``Sentiment polarity
  detection for software development,'' \emph{Empirical Software Engineering},
  vol.~23, no.~3, pp. 1352--1382, 2018.

\bibitem{lin2018sentiment}
B.~Lin, F.~Zampetti, G.~Bavota, M.~Di~Penta, M.~Lanza, and R.~Oliveto,
  ``Sentiment analysis for software engineering: How far can we go?'' in
  \emph{Proceedings of 40th International Conference on Software Engineering},
  2018.

\bibitem{felbo2017using}
B.~Felbo, A.~Mislove, A.~S{\o}gaard, I.~Rahwan, and S.~Lehmann, ``Using
  millions of emoji occurrences to learn any-domain representations for
  detecting sentiment, emotion and sarcasm,'' \emph{arXiv preprint
  arXiv:1708.00524}, 2017.

\bibitem{imtiaz2018sentiment}
N.~Imtiaz, J.~Middleton, P.~Girouard, and E.~Murphy-Hill, ``Sentiment and
  politeness analysis tools on developer discussions are unreliable, but so are
  people,'' 2018.

\bibitem{boia2013worth}
M.~Boia, B.~Faltings, C.-C. Musat, and P.~Pu, ``A:) is worth a thousand words:
  How people attach sentiment to emoticons and words in tweets,'' in
  \emph{Social computing (socialcom), 2013 international conference on}.\hskip
  1em plus 0.5em minus 0.4em\relax IEEE, 2013, pp. 345--350.

\bibitem{park2013emoticon}
J.~Park, V.~Barash, C.~Fink, and M.~Cha, ``Emoticon style: Interpreting
  differences in emoticons across cultures.'' in \emph{ICWSM}, 2013.

\bibitem{marvin1995spoof}
L.-E. Marvin, ``Spoof, spam, lurk, and lag: The aesthetics of text-based
  virtual realities,'' \emph{Journal of Computer-Mediated Communication},
  vol.~1, no.~2, p. JCMC122, 1995.

\bibitem{rezabek1998visual}
L.~Rezabek and J.~Cochenour, ``Visual cues in computer-mediated communication:
  Supplementing text with emoticons,'' \emph{Journal of Visual Literacy},
  vol.~18, no.~2, pp. 201--215, 1998.

\bibitem{witmer1997line}
D.~F. Witmer and S.~L. Katzman, ``On-line smiles: Does gender make a difference
  in the use of graphic accents?'' \emph{Journal of Computer-mediated
  communication}, vol.~2, no.~4, p. JCMC244, 1997.

\bibitem{hogenboom2013exploiting}
A.~Hogenboom, D.~Bal, F.~Frasincar, M.~Bal, F.~de~Jong, and U.~Kaymak,
  ``Exploiting emoticons in sentiment analysis,'' in \emph{Proceedings of the
  28th Annual ACM Symposium on Applied Computing}.\hskip 1em plus 0.5em minus
  0.4em\relax ACM, 2013, pp. 703--710.

\bibitem{bahri2018novel}
S.~Bahri, P.~Bahri, and S.~Lal, ``A novel approach of sentiment classification
  using emoticons,'' \emph{Procedia Computer Science}, vol. 132, pp. 669--678,
  2018.

\bibitem{novak2015sentiment}
P.~K. Novak, J.~Smailovi{\'c}, B.~Sluban, and I.~Mozeti{\v{c}}, ``Sentiment of
  emojis,'' \emph{PloS one}, vol.~10, no.~12, p. e0144296, 2015.

\bibitem{hogenboom2015exploiting}
A.~Hogenboom, D.~Bal, F.~Frasincar, M.~Bal, F.~De~Jong, and U.~Kaymak,
  ``Exploiting emoticons in polarity classification of text.'' \emph{J. Web
  Eng.}, vol.~14, no. 1\&2, pp. 22--40, 2015.

\bibitem{liu2012emoticon}
K.-L. Liu, W.-J. Li, and M.~Guo, ``Emoticon smoothed language models for
  twitter sentiment analysis.'' in \emph{Aaai}, 2012.

\bibitem{davidov2010enhanced}
D.~Davidov, O.~Tsur, and A.~Rappoport, ``Enhanced sentiment learning using
  twitter hashtags and smileys,'' in \emph{Proceedings of the 23rd
  international conference on computational linguistics: posters}.\hskip 1em
  plus 0.5em minus 0.4em\relax Association for Computational Linguistics, 2010,
  pp. 241--249.

\bibitem{wang2015sentiment}
H.~Wang and J.~A. Castanon, ``Sentiment expression via emoticons on social
  media,'' in \emph{Big Data (Big Data), 2015 IEEE International Conference
  on}.\hskip 1em plus 0.5em minus 0.4em\relax IEEE, 2015, pp. 2404--2408.

\bibitem{read2005using}
J.~Read, ``Using emoticons to reduce dependency in machine learning techniques
  for sentiment classification,'' in \emph{Proceedings of the ACL student
  research workshop}.\hskip 1em plus 0.5em minus 0.4em\relax Association for
  Computational Linguistics, 2005, pp. 43--48.

\bibitem{wolf2000emotional}
A.~Wolf, ``Emotional expression online: Gender differences in emoticon use,''
  \emph{CyberPsychology \& Behavior}, vol.~3, no.~5, pp. 827--833, 2000.

\bibitem{miller2016blissfully}
H.~Miller, J.~Thebault-Spieker, S.~Chang, I.~Johnson, L.~Terveen, and B.~Hecht,
  ``Blissfully happy” or “ready to fight”: Varying interpretations of
  emoji,'' \emph{Proceedings of ICWSM}, vol. 2016, 2016.

\bibitem{DuenasPerceval}
\BIBentryALTinterwordspacing
S.~Due\~{n}as, V.~Cosentino, G.~Robles, and J.~M. Gonzalez-Barahona,
  ``Perceval: Software project data at your will,'' in \emph{Proceedings of the
  40th International Conference on Software Engineering: Companion
  Proceeedings}, ser. ICSE '18.\hskip 1em plus 0.5em minus 0.4em\relax New
  York, NY, USA: ACM, 2018, pp. 1--4. [Online]. Available:
  \url{http://doi.acm.org/10.1145/3183440.3183475}
\BIBentrySTDinterwordspacing

\bibitem{mantyla2018nlon}
M.~V. M{\"a}ntyl{\"a}, F.~Calefato, and M.~Claes, ``Natural language or not
  (nlon)-a package for software engineering text analysis pipeline,''
  \emph{arXiv preprint arXiv:1803.07292}, 2018.

\bibitem{plutchik1991emotions}
\emph{The emotions}.\hskip 1em plus 0.5em minus 0.4em\relax University Press of
  America, 1991.

\end{thebibliography}

\end{document}